# The Environments of Fast Radio Bursts Viewed Using Adaptive Optics

Michele N. Woodland[1] [ID], Alexandra G. Mannings[2], J. Xavier Prochaska[2,3,4] [ID], Stuart D. Ryder[5,6] [ID], Lachlan Marnoch[5,6] [ID], Regina A. Jorgenson[1] [ID], Sunil Simha[2] [ID], Nicolas Tejos[7] [ID], Alexa Gordon[8] [ID], Wen-fai Fong[8] [ID], Charles D. Kilpatrick[8] [ID], Adam T. Deller[9] [ID], and Marcin Glowacki[10] [ID]

[1] Maria Mitchell Observatory, Nantucket, MA 02554, USA
[2] Department of Astronomy and Astrophysics, University of California, Santa Cruz, CA 95064, USA
[3] Kavli Institute for the Physics and Mathematics of the Universe (Kavli IPMU), 5-1-5 Kashiwanoha, Kashiwa, 277-8583, Japan
[4] Division of Science, National Astronomical Observatory of Japan, 2-21-1 Osawa, Mitaka, Tokyo 181-8588, Japan
[5] School of Mathematical and Physical Sciences, Macquarie University, NSW 2109, Australia
[6] Astrophysics and Space Technologies Research Centre, Macquarie University, Sydney, NSW 2109, Australia
[7] Instituto de Física, Pontificia Universidad Católica de Valparaíso, Casilla 4059, Valparaíso, Chile
[8] Center for Interdisciplinary Exploration and Research in Astrophysics (CIERA) and Department of Physics and Astronomy, Northwestern University, Evanston, IL 60208, USA
[9] Centre for Astrophysics and Supercomputing, Swinburne University of Technology, Hawthorn, VIC 3122, Australia
[10] International Centre for Radio Astronomy Research (ICRAR), Curtin University, Bentley, WA 6102, Australia

## Abstract

We present observations from the Gemini Multi-Conjugate Adaptive Optics System/Gemini South Adaptive Optics Imager at Gemini South of five fast radio burst (FRB) host galaxies of FRBs with subarcsecond localizations. We examine and quantify the spatial distributions and locations of the FRBs with respect to their host galaxy light distributions, finding a median host-normalized offset of 2.09 half-light radii ($r_e$) and the trend that these FRBs occur in fainter regions of their host galaxies. When combined with the FRB host galaxy sample from Mannings et al., we find that FRBs are statistically distinct from Ca-rich transients in terms of light at the source location and from SGRBs and LGRBs in terms of host-normalized offset. We further find that most FRBs are in regions of elevated local stellar mass surface densities in comparison to the mean global values of their hosts. This, along with the finding that the FRB locations trace the distribution of stellar mass, points toward a possible similarity of the environments of CCSNe and FRBs. We also find that four out of five FRB hosts exhibit distinct spiral arm features, and the bursts originating from such hosts tend to appear on or close to their host's spiral structure, with a median distance of $0.53 \pm 0.27$ kpc. With many well-localized FRB detections looming on the horizon, we will be able to better characterize the properties of FRB environments relative to their host galaxies and other transient classes. Such insights may only require us to double the number of FRBs with subarcsecond localizations.

*Unified Astronomy Thesaurus concepts:* Radio transient sources (2008); Galaxy photometry (611)

## 1. Introduction

Fast radio bursts (FRBs) are luminous millisecond-duration bursts of radio waves that originate from extragalactic sources. FRBs were first discovered in 2007 (Lorimer et al. 2007), but it was not until 2017 that the FRB—FRB 20121102A—was precisely localized and associated with a host galaxy at $z = 0.19$, providing the first direct evidence of their cosmological origins (Chatterjee et al. 2017; Marcote et al. 2017; Tendulkar et al. 2017). However, despite the continued FRB detections, and subsequent host galaxy studies (e.g., Tendulkar et al. 2017; Heintz et al. 2020; CHIME/FRB Collaboration et al. 2021; Chittidi et al. 2021; Kirsten et al. 2022; Bhandari et al. 2023; Law et al. 2023; Ryder et al. 2023), the sources of FRBs remain largely a mystery.

While some FRBs are known to repeatedly burst, most FRBs are only associated with a single burst and are considered apparent nonrepeaters (e.g., CHIME/FRB Collaboration et al. 2021). Repetition aids in localization, as demonstrated by the fact that all of the first precisely localized FRBs were repeaters until the Commensal Real-time ASKAP Fast Transient (CRAFT) survey (e.g., Bannister et al. 2017; Marcote et al. 2017; Bhandari et al. 2020a; Marcote et al. 2020; Tendulkar et al. 2020). However, with advances in FRB searches more broadly, the capability to localize apparent nonrepeaters to subarcsecond scales is sharply increasing. A greater number of FRBs are now being detected with subarcsecond and even milliarcsecond localizations by fast transient searches using radio interferometers (Driessen et al. 2023; Law et al. 2023). A high volume of well-localized FRB detections (multiple per day) from the CHIME Outriggers, CRACO (successor to CRAFT) on Australian Square Kilometre Array Pathfinder (ASKAP), and DSA-110 looms on the horizon (Kocz et al. 2019; Leung et al. 2021).

Though an FRB-like burst was localized to a Galactic magnetar (SGR 1935+2154; Bochenek et al. 2020; CHIME/FRB Collaboration et al. 2020), a great diversity of host galaxy characteristics and FRB locations could support the efficacy of multiple progenitor pathways. FRBs have been localized to a globular cluster (Bhardwaj et al. 2021; Kirsten et al. 2022), star-forming dwarf galaxies (Marcote et al. 2017; Bhandari et al. 2023), and massive spiral galaxies with varying offsets and associations to underlying morphological structure (e.g., Marcote et al. 2020; Chittidi et al. 2021; Mannings et al. 2021;

Dong et al. 2023). The repeating bursts FRBs 20121102A and 20190520B have also been associated with compact, persistent radio sources (Marcote et al. 2017; Niu et al. 2022). Thus, more well-localized FRB detections and host observations are necessary to create a larger sample from which we can derive meaningful statistics and continue making robust associations.

Local environment studies, in conjunction with the characterization of global host properties and constraints on multiwavelength or persistent counterparts, have been transformative in our understanding of the origins of other transients such as short- and long-duration gamma-ray bursts (e.g., Fruchter et al. 2006; Fong & Berger 2013; Fong et al. 2017; Zhang et al. 2018) and various types of supernovae (e.g., Kuncarayakti et al. 2018; Hsu et al. 2023). In this study, we utilize adaptive optics (AO) imaging from the Gemini South telescope to study the local environments of five FRBs along with their global host properties to help better understand FRBs and their origins.

Improvements in FRB detection rates and localization precision will have—and are already having—important implications for progenitor science. A key to understanding FRB sources lies in our ability to characterize the local environments of these bursts. This, in combination with analysis of burst properties and propagation effects such as dispersion measures, time variability of rotation measures, and scattering can provide key insights to FRB production mechanisms and sources. Many such local environment studies have relied on space-based observations to achieve depth and resolution that reach subarcsecond scales (Mannings et al. 2021, etc.). Unfortunately, as the sample of FRBs grows, it may not be feasible to follow up each event with space-based resources. Therefore, we find it necessary to develop a ground-based follow-up approach that utilizes AO as an alternative for accommodating high-spatial resolution for high-volume follow-up efforts. While it is true that more information about burst properties can provide much-needed information, we use this paper to focus on the investigation of the efficacy of ground-based AO observations in relation to follow-up work. Future papers will explore these details, possible correlations, and therefore physical explanations of observed effects.

This paper is organized as follows: in Section 2, we describe the sample selection and observations. In Section 3, we present our analysis and results on the brightness of the host galaxy at the FRB location in comparison with the rest of the galaxy. In Section 4, we discuss the FRB physical locations and offsets relative to the structure of their host galaxies. We discuss these results in Section 5 and conclude in Section 6. Throughout the paper, we employ a Planck cosmology with $H_0 = 67.8 \text{ km s}^{-1} \text{ Mpc}^{-1}$, $\Omega_M = 0.308$, and $\Omega_\Lambda = 0.692$ (Planck Collaboration et al. 2016). However, it can be noted that changes in these values do not significantly affect the results.

## 2. FRB Data and Sample Selection

### 2.1. Sample of FRB Host Galaxies

Here we present observations of seven FRB host galaxies obtained with the Gemini South Adaptive Optics Imager (GSAOI) and the Gemini Multi-Conjugate Adaptive Optics System (GeMS) on Gemini South. The data for these FRB host galaxies were collected between 2021 July and 2022 April as part of the programs GS-2021A-C-2, GS-2021B-C-3, and GS-2022A-C-2.

All of the host galaxies in our sample have spectroscopically confirmed redshifts (Bhandari et al. 2020a; Heintz et al. 2020; Bhandari et al. 2023; Glowacki et al. 2023), and their FRBs were discovered with the ASKAP telescope through the CRAFT survey. These are considered secure associations with PATH (Aggarwal et al. 2021) posterior probabilities of $P(O|x) \geqslant 0.93$. None of the FRB samples in this sample have yet been observed to repeat. Many of these FRBs have been presented in previous works (e.g., Prochaska et al. 2019; Heintz et al. 2020; Mannings et al. 2021; Bhandari et al. 2023; Glowacki et al. 2023), but the hosts of FRBs 20210807D and 20211212A are newly presented here. In a companion paper, we describe imaging of the hosts of FRBs 20210807D and 20211212A, including the FRB localization, using the Very Large Telescope (A. Deller et al., in preparation). To be included in this sample, the hosts also had to be in a field that satisfied the guide star requirements outlined in Section 2.2.

### 2.2. Host Observations

Laser-guide-star-AO-assisted near-IR observations of the FRB host galaxies were obtained with GeMS/GSAOI on the 8.1 m Gemini South telescope. GSAOI is fed by GeMS and covers an $85'' \times 85''$ field of view (FoV) with a pixel scale of $0\rlap{.}''0197 \text{ pixel}^{-1}$, delivering close to diffraction-limited images between 0.9 and 2.4 $\mu$m. Uniform AO correction across the full GeMS FoV requires up to three natural guide stars (NGSs) in addition to the five-point sodium laser guide star (LGS) pattern. However, partial AO correction is still possible with two (or even one) NGSs of sufficient brightness ($m_R < 15.5$ mag) within the $1'$ patrol field of the wave front sensor probes, as well as one on-detector guide window star ($m_H < 13.5$ mag) within the $40''$ FoV of any of the four GSAOI detectors at all dither positions.

Each target was imaged in the $K_s$ filter [1962.58–2335.92 nm][11] using a nine-step dither pattern, with two coadds of 60 s at each position and a step size large enough (5″) to cover the gaps between the detectors. The targets were positioned within one of GSAOI's four detectors, with the array orientation set by the locations of the NGS. Including overheads and interruptions due to aircraft and satellite avoidance, a complete observation required 1–1.5 hr. These observations are summarized in Table 1.

For FRB 20210117A, the host was not detected with GeMS/GSAOI at the FRB coordinates and only five out of nine dithers were completed. This host is a faint dwarf galaxy as discussed by Bhandari et al. (2023), as is consistent with our nondetection.

### 2.3. Image Processing

The data were retrieved from the Gemini Observatory Archive (GOA) at NSF's NOIRLab. The data were reduced using the Gemini Data Reduction for Astronomy from Gemini Observatory North and South (Labrie et al. 2019) package version 3.1.

The dark current is very low for GSAOI ($\sim$0.01 e- s$^{-1}$ pix$^{-1}$); thus dark subtraction was not necessary. A master flat field was created from a series of lamp-on and lamp-off exposures. A stack was made for each, and the lamp-off stack was then subtracted from the lamp-on stack. This result was normalized.

---

[11] https://www.gemini.edu/instrumentation/gsaoi/components

**Table 1**
Host Properties

| FRB | R.A._Host (J2000) | Decl._Host (J2000) | $\sigma_{Host}$ (arcsec) | $z$ | $M_\star$ ($10^9 M_\odot$) | NGS | Date (UT) | Exposure Time (s) |
|---|---|---|---|---|---|---|---|---|
| 20180924B | 21h44m25$^s$2 | −40d54m00$''$9 | 0.13 | 0.321 | 24.5 | 3 | 2021 Aug 1 | 540 |
| 20181112A | 21h49m23$^s$7 | −52d58m15$''$3 | 0.13 | 0.475 | 7.4 | 1 | 2021 Aug 2 | 480 |
| 20191001A | 21h33m24$^s$4 | −54d44m54$''$3 | 0.29 | 0.234 | 53.7 | 2 | 2021 Aug 1 | 1860 |
| 20210117A | 22h39m55$^s$1 | −16d09m05$''$4 | ⋯ | 0.214 | 0.4 | 1 | 2021 Aug 2 | 240 |
| 20210807D | 19h56m52$^s$8 | −00d45m44$''$5 | 1.03 | 0.129 | 93.3 | 3 | 2021 Sep 13 | 1020 |
| 20211127I | 13h19m13$^s$9 | −18d50m16$''$1 | 0.28 | 0.047 | 24.5 | 3 | 2022 Apr 15 | 2040 |
| 20211212A | 10h29m24$^s$2 | +01d21m39$''$0 | 0.71 | 0.071 | 19.0 | 1 | 2022 Apr 15 | 1140 |

**Note.** R.A._Host and decl._Host are the host coordinates in the GeMS/GSAOI image WCS. $\sigma_{Host}$ was calculated by summing the R.A. and decl. errors in quadrature. For FRB 20181112A, a $\sigma_{Host}$ represents the uncertainty in the host position due to astrometric corrections. For FRB 20210117A, no host was detected, and the astrometry of the image was not corrected; thus we do not have a value for $\sigma_{Host}$. NGS is the number of natural guide stars used during observation. Date refers to the date of observation.

The static bad pixel masks (BPMs) were fetched from the GOA. The science files were then flat field corrected, aligned, and stacked. The final reduced images are shown in Figure 1.

### 2.4. Astrometry and Uncertainties

We performed absolute astrometry using the Gaia DR2 and Gaia DR3 catalogs. To do this, we utilized PHOTUTILS. SEGMENTATION to create a segmentation map of each respective image. We then cross-matched these $x$ and $y$ centroids with the Gaia sources within 3$''$ of these coordinates. For the cases where the world coordinate system (WCS) information of the image is offset by more than 3$''$, a manual correction was done by cross-matching the $x$ and $y$ centroid values with the determined Gaia sources based upon a visual (by-eye) association. A new WCS header was then created with this information. We calculated the rms uncertainty of the astrometry of the images as $\sigma_{ast}$.

Due to lack of Gaia sources in the fields for FRBs 20181112A and 20211212A, the process above was followed but using the Dark Energy Survey (Abbott et al. 2021) and Panoramic Survey Telescope and Rapid Response System (Chambers et al. 2016; Flewelling et al. 2020), respectively, as reference catalogs from which to extract stellar catalogs.

We then used GALIGHT (Ding et al. 2021) to determine the host galaxy centroid positions and their associated uncertainties ($\sigma_{host}$). Inputs for GALIGHT included the radius for cutout around the galaxy and guesses for the galaxy centroid position and position angle. The positions and $\sigma_{host}$ values are listed in Table 1.

Since we analyze the host brightness at the FRB location, the associated localization uncertainty ($\sigma_{FRB}$) is also incorporated into our calculations. The statistical and systematic uncertainties from the FRB localizations are included in $\sigma_{FRB}$ (Table 2).

## 3. Light at FRB Locations

Placing transients in relation to the distribution of various wavelengths of light (related to ionized gas, neutral gas, stellar components, etc.) in their hosts can be an important piece of evidence for tying them to a source or progenitor environment (Fruchter et al. 2006; Fong et al. 2010; Fong & Berger 2013). In the infrared, we are most sensitive to the distribution of the older (more numerous) stellar populations. Here we complete morphological studies of each host and study the light at the location of the burst. In combination with one another, these analyses provide information on the local environment of the burst and therefore likely sources. For FRB 20191001A, it was found that the coordinates published previously in other works were offset in the incorrect direction in decl. (see Bhandari et al. 2020b). The coordinates presented in this paper are the updated coordinates.

### 3.1. Galaxy Light Profile Fitting

We use GALIGHT (Ding et al. 2021) to fit the light profile for five of the FRB hosts to determine the half-light radii ($r_e$) and study low surface brightness structures and host morphologies. Owing to the faintness of the hosts for FRBs 20181112A and 20210117A, we were unsuccessful at creating a segmentation image with GALIGHT and could not complete related analyses. These analyses are also limited by the relatively poor localization precision of FRB 20181112A, where any relationship to the underlying substructure would be difficult to determine and, therefore, interpret.

Using GALIGHT, we first created a cutout around the host galaxy and masked any unwanted sources. We then defined the coordinates of at least four stars in the original image and created point-spread function (PSF) models to choose from and utilize. We used the GALIGHT model to compute residuals from Sérsic profile fits, where the Sérsic index $n$ defines the steepness of the radial surface brightness profile. A larger $n$ value indicates a steeper inner profile and an extended outer wing. The hosts of FRB 20180924B, FRB 20211127I, and FRB 20211212A were fit using two Sérsic components, equating to a disk and a bulge. FRBs 20191001A and 20210807D were fit utilizing three Sérsic components.

The GALIGHT half-light radii $r_e$ are presented in Table 3. The values correspond to the effective radius of the disk component of the fit produced by GALIGHT. The $r_e$ values range from 0$''$43 to 2$''$80. We then converted these angular offsets to physical offsets using the host redshifts and Planck 15 cosmology as detailed in Section 1. The physical effective radii range from 2.07 to 6.67 kpc, those values belonging to the host galaxies for FRB 20180924B and FRB 20191001A, respectively. Inclination is accounted for.

The residual images from the GALIGHT fits are shown in Figure 2. After removing the smooth light of the galaxies from the GALIGHT Sérsic models, the spiral structure is quite apparent for four out of five hosts, with hints of spiral structure for the host of FRB 20180924B. This host provides a test case for comparisons between space- and ground-based imaging. Though the Hubble Space Telescope (HST) image (and

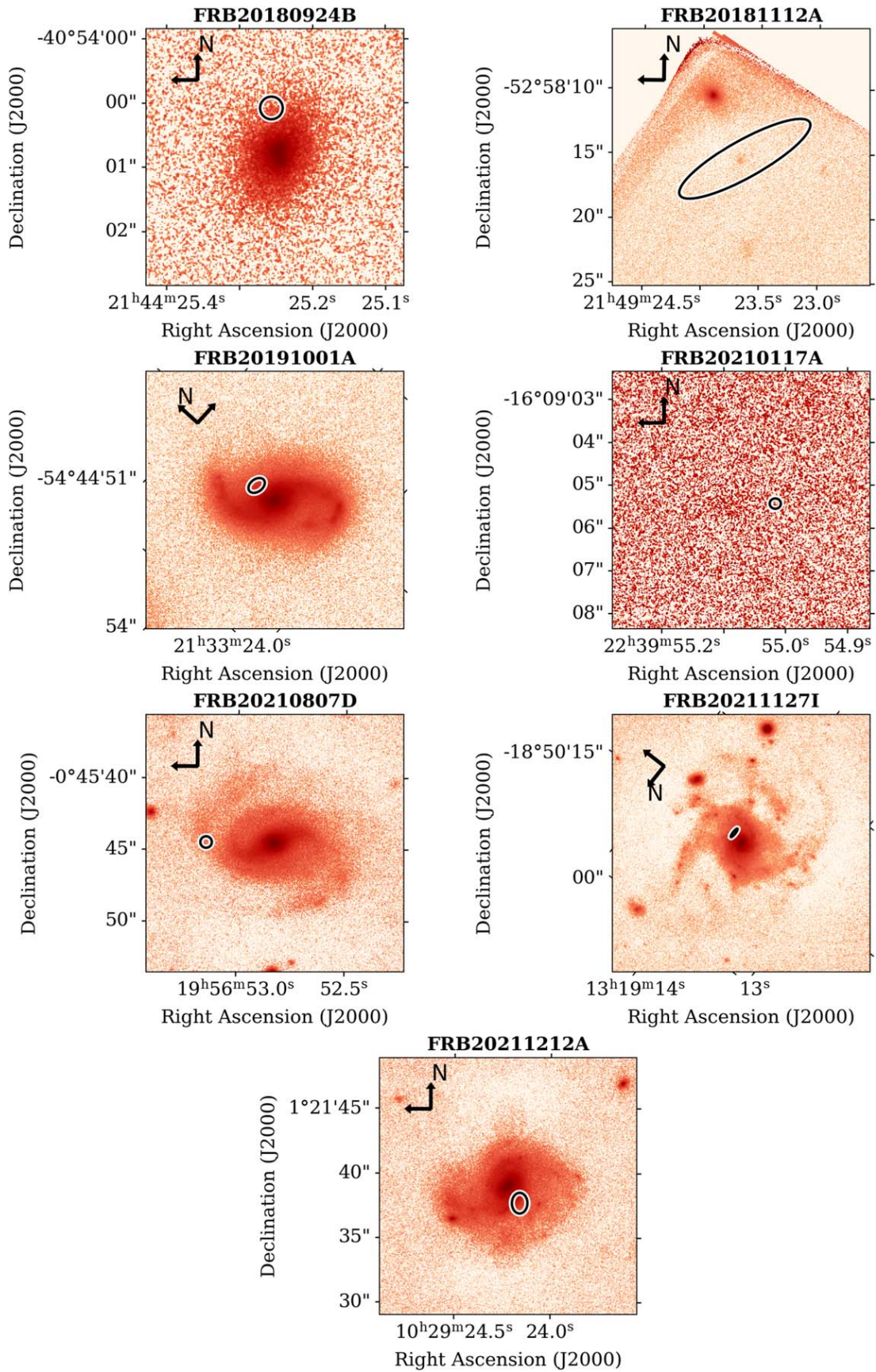

**Figure 1.** GeMS/GSAOI imaging of the seven host galaxies in our sample in the $K_s$ filter. The black ellipse in each image represents the 68% confidence level FRB localization region. The black arrows indicate north and east.

**Table 2**
FRB Positional Uncertainties

| FRB | RA$_{FRB}$ (J2000) | Dec$_{FRB}$ (J2000) | $a$ (arcsec) | $b$ (arcsec) | PA (deg) | Reference |
|---|---|---|---|---|---|---|
| 20180924B | 21h44m25ˢ25 | −40d54m00ˢ10 | 0.18 | 0.18 | 0 | (1) |
| 20181112A | 21h49m23ˢ62 | −52d58m15ˢ39 | 5.75 | 1.44 | 120 | (2) |
| 20191001A | 21h33m24ˢ42 | −54d44m15ˢ58 | 0.27 | 0.19 | 79 | (3) |
| 20210117A | 22h39m55ˢ01 | −16d09m05ˢ45 | 0.13 | 0.12 | 90 | (4) |
| 20210807D | 19h56m53ˢ14 | −00d45m44ˢ49 | 0.40 | 0.40 | 0 | (5), (6) |
| 20211127I | 13h19m14ˢ08 | −18d50m16ˢ69 | 0.80 | 0.20 | 0 | (5), (6) |
| 20211212A | 10h29m24ˢ16 | +01d21m37ˢ67 | 0.80 | 0.80 | 0 | (5), (6) |

**Note.** $a$ refers to the angular size of the semimajor axis describing the $1\sigma$ statistical ($a_{stat}$) and systematic ($a_{sys}$) uncertainties added in quadrature. The same applies to $b$, the semiminor axis. PA refers to the position angle of the error ellipse on the sky. For FRB 20191001A, it was found that the coordinates published previously in other works were offset in the incorrect direction in decl. (see Bhandari et al. 2020b). The coordinates presented here are the updated coordinates.

**References.** (1) Bannister et al. (2019); (2) Prochaska et al. (2019); (3) Bhandari et al. (2020b); (4) Bhandari et al. (2020a); (5) Gordon et al. (2023); (6) A. Deller et al., in preparation.

associated residual) presented in Mannings et al. (2021) showed more extended, spiral structure, GeMS+GSAOI does somewhat recover the structure in this host as evidenced by Figure 2. The spiral structure for FRB 20211212A is newly presented here with further discussion in A. Deller et al. (in preparation). The spiral structure for FRB 20180924B was previously known (Mannings et al. 2021), as well as the structures for FRB 20191001A (Bhandari et al. 2020b), FRB 20210807D, and FRB 20211127I (Glowacki et al. 2023).

While the spiral structure is already apparent in Figure 1, the arms are more defined in the residual images and thus better for the spiral arm offset analysis described in Section 4.2. For each host, the FRB occurs either on or close to a spiral arm of the galaxy. The host of FRB 20211127I is now one of only four reported FRBs with a clear bar structure observed in the host galaxy (along with the hosts of FRB 20190608B, FRB 20201124A, and FRB 20220319D; Chittidi et al. 2021; Xu et al. 2022; Ravi et al. 2023).

### 3.2. Stellar Mass Surface Density

Due to the high-precision subarcsecond localizations of the FRBs in this sample as matched by the image resolution, photometric measurements could be made at the immediate burst position. The following photometric analysis was modeled after that performed by Mannings et al. (2021) on HST imaging in service of better understanding the local environment properties and how they compare to global host properties.

We measured the stellar mass surface density ($\Sigma_{M*}$) at the location of the FRB by determining the IR surface brightness within the FRB localization region. We account for the uncertainty in the FRB localization in our estimate. We do so by computing a weighted average where the relative weights are given by a 2D Gaussian probability distribution function (PDF) along the axes of the localization uncertainty. We create circular apertures with a radius of 3 pixels, equating to the size of the PSF, at each pixel within the FRB localization. These aperture sums are then weighted accordingly, with the resolution of the Gaussian grid and binning being limited by the image pixel scale. We then divide this weighted average by the area of the aperture to get an aperture sum per square arcsecond, or surface brightness.

We separately determine the magnitude of the host galaxies using the GSAOI zero-point and the corresponding flux. The magnitude of each host galaxy is presented in Table 3. We then calculate the ratio between the flux at the FRB site to the flux of the entire galaxy and adopt the total stellar mass estimates from Heintz et al. (2020) and Bhandari et al. (2020a) to estimate the host $\Sigma_{M*}$ (Table 3). The relationship between the local stellar mass surface density at the FRB location and the average stellar mass surface density of the entire galaxy is shown in Figure 3.

Figure 3 reveals that the stellar mass surface densities $\Sigma_{M*}$ at the FRB locations are elevated compared to the global stellar mass surface densities of their host galaxies for four out of five hosts. The other host (FRB 20180924B) has a stellar mass surface density lower at the FRB location in comparison to the galaxy as a whole; however, this value is fairly close to the 1:1 line. Error bars for each of the points are significantly smaller than the marker size. We use Milky Way giant molecular clouds (GMCs; Lada & Dame 2020) as a means of comparing the locations of FRBs to sites of active star formation within the Milky Way. The stellar mass surface density at the FRB positions relative to their hosts is 3 times or more than that shown for the Galactic GMCs, which is $\approx 35 \times 10^6\ M_\odot\ kpc^{-2}$ according to Lada & Dame (2020).

### 3.3. Luminosity Constraints on Background Galaxies

With hosts such as those of FRBs 20121102A (Tendulkar et al. 2017), 20190711A (Heintz et al. 2020), and 20210117A (Bhandari et al. 2023), associations show that FRBs can originate in less massive—and therefore fainter—hosts than the primarily massive, star-forming hosts shown in this sample. With our high-precision, deep AO imaging, we may test the scenario in which the true host is a dwarf and possible companion to the putative host, along with the scenario where a higher-$z$ background galaxy may be the true host.

To put a constraint on a fainter host galaxy candidate at the FRB position, we use the GALIGHT residual images as shown in Figure 2, where the smooth elliptical components from the putative host have been removed to derive point-source limiting magnitudes (shown in Table 3) at the FRB position. We calculate the residual flux using a circular aperture of 0″5 in diameter and compute the net standard deviation for the pixels within this aperture. We then take the flux measurement with 5 times the net standard deviation as the upper limit on any point-source flux that could be detected from the residual images. Our limits and visual inspection of the images indicate that there are no galaxies brighter than an apparent magnitude

Table 3
Derived Properties of Host Galaxies

| FRB | $r_e$ (arcsec) | $r_e$ (kpc) | $i$ (deg) | Host Magnitude (AB mag) | Limiting Magnitude (AB mag) | $\Sigma_{M_*\text{FRB}}$ ($10^8\,M_\odot\,\text{kpc}^{-2}$) |
|---|---|---|---|---|---|---|
| 20180924B | 0.43 ± 0.01 | 2.07 ± 0.03 | 51 | 18.841 ± 0.006 | 28.0 | 1.832 ± 0.001 |
| 20191001A | 1.74 ± 0.01 | 6.67 ± 0.04 | 59 | 15.883 ± 0.003 | 26.4 | 3.278 ± 0.001 |
| 20210807D | 2.76 ± 0.01 | 6.56 ± 0.04 | 60 | 16.729 ± 0.003 | 27.1 | 1.695 ± 0.001 |
| 20211127I | 2.80 ± 0.01 | 2.67 ± 0.01 | 33 | 16.059 ± 0.002 | 28.0 | 4.612 ± 0.004 |
| 20211212A | 2.36 ± 0.01 | 3.29 ± 0.01 | 40 | 15.790 ± 0.002 | 25.8 | 2.461 ± 0.001 |

**Note.** FRB 20181112A and FRB 20210117A are not included in these calculations due to the large FRB positional uncertainty relative to the size of the host and the detection not being significant enough for the host to be identified by GALIGHT (see Section 3.1). Limiting magnitudes correspond to $5\sigma$ detection thresholds.

of 26 at the FRB position. At these redshifts, the limits of $L_{IR} \lesssim (0.02–0.39) \times 10^6\,L_\odot$ are significantly deeper than the luminosity of even the faintest known dwarf galaxies associated with FRBs, namely $10^8 < L_\odot < 10^9$ (e.g., Bassa et al. 2017; Bhandari et al. 2023).

### 3.4. Fractional Flux

We next examine the location of the FRBs relative to their host galaxy light distributions, a measure known as fractional flux ($F_F$). The $F_F$ is independent of host size and morphology, as the measurement determines the fraction of host light fainter than the flux at the burst position. A value of 1 would indicate that the transient occurred in the brightest location within the host. This measure follows the $F_F$ calculation used in Fruchter et al. (2006), which was an important factor in differentiating long gamma-ray bursts (LGRBs) from core-collapse supernovae (CCSNe) locations in their respective hosts. In the IR, as with the $K$-band images presented here, we are able to probe the relationship between FRBs and the distribution of stellar mass in their hosts. A higher fractional flux in this regime points toward FRB locations coincident with high concentrations of stellar mass–like the centers of galaxies.

To determine $F_F$, we create a 2D cutout around each host galaxy large enough to include background pixels. We then identify the $N$ pixels that lie within the $3\sigma$ FRB localization ellipse. The fractional flux for a given ($i$th) pixel within the localization ellipse is given by

$$F_{F,i} = \frac{\Sigma_j^N (F_j < F_i)}{\Sigma_j^N F_j},\qquad(1)$$

where $F_i$ is the flux value in pixel $i$. We then weight the $F_{F,i}$ values by the 2D Gaussian distribution of the FRB localization ellipse. The $F_F$ values for each FRB can be found in Table 4. The values range from 0.24 to 0.64 with a median of 0.59. The lower and upper limits were set by FRBs 20211127I and 20211212A, respectively.

We combine our sample with the Mannings et al. (2021) sample to determine the distribution of these values for a larger sample of ten FRB host galaxies. The additional hosts that were chosen those with WFC3/IR F160W imaging that were not included in the sample imaged with Gemini (FRB 20121102A, FRB 20190102C, FRB 20190608B, FRB 20190711A, and FRB 20190714A). This will hereafter be referred to as the combined FRB sample, whereas the GeMS/GSAOI sample will be referred to as such.

In order to determine the uncertainty on the cumulative distributions, we utilize the method of Heintz et al. (2020) to create 10,000 realizations of asymmetric Gaussian PDFs using the calculated offset errors for each FRB. Finally, we use a bootstrap method to compute a cumulative distribution function (CDF) of the bootstrapped sample and calculate the median of these CDFs, in addition to the lower and upper bounds for each bin. Figure 4 shows the fractional flux CDF for the combined FRB sample along with the cumulative distributions of other transient class samples pulled from other publications (Fong et al. 2010; Fong & Berger 2013; Wang et al. 2013; Lunnan et al. 2015; Blanchard et al. 2016).

In Figure 4 we can see that the FRB cumulative distribution (shown in black) differs markedly from the distributions of short gamma-ray bursts (SGRBs, red) and LGRBs (orange). The FRB distribution appears similar to that of the 1-1 line (blue), Type Ia supernovae (SNe, purple) and CCSNe (green). However, due to the small sample size (10 FRBs) and the FRB positional uncertainties, the overall uncertainty is substantial and cannot be ignored. We also assume here that the ten FRBs have the same physical origin, which cannot necessarily be taken for granted. We explore these relationships with more rigorous statistical testing.

Using the Kolmogorov–Smirnov (KS) test, we test the null hypothesis that FRBs come from the same underlying distribution as the samples of other transient classes shown here. We require $P_{KS} < 0.05$ in order to reject this null hypothesis. We test against CCSNe, SGRBs, LGRBs, and Type Ia SNe. The resulting $P_{KS}$ values are shown in the legend of Figure 4. We find that the KS test rejects the null hypothesis that SGRBs or LGRBs are from the same underlying population as FRBs. CCSNe and Type Ia SNe could not be rejected.

## 4. FRB Physical Locations and Offsets

In this section, we study the locations of the FRBs in our sample with respect to their host galaxy centers and spiral structure. We introduce their angular and physical offsets ($\theta$ and $\delta R$, respectively); their "host-normalized" offsets ($\delta R/r_e$), which are normalized by the half-light radii $r_e$ of their host galaxies; and the locations with respect to the closest spiral arm of their host galaxy (spiral offset, SO). We then compare the galactocentric offset distributions to distributions of other transient events as we did for $F_F$ in Section 3.4. We also investigate the relationship between FRBs and spiral arms and how this compares to other transient populations and previous FRB studies.

### 4.1. Galactocentric Offsets

The shape, size, and orientation of FRB localization ellipses must be taken into account when determining the angular,

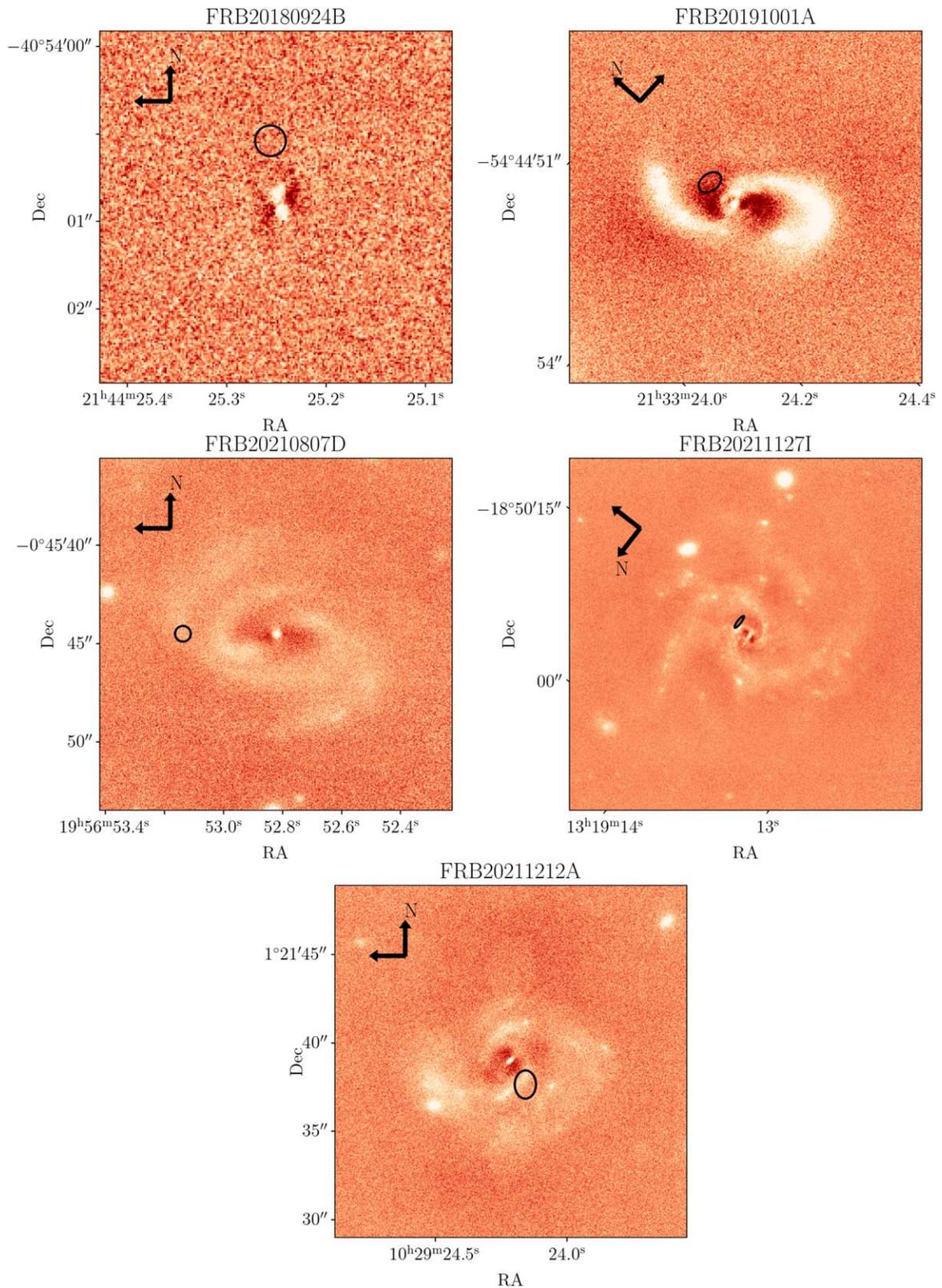

**Figure 2.** Residual images produced by subtracting the GALIGHT model from the original images for the objects with host detections. The black ellipse in each image represents the FRB localization region. North and east are indicated by the black arrows in the top left. Light regions show a flux excess, while dark regions show a flux deficit.

physical, and host-normalized offsets from the center of their host galaxies. We define ellipsoidal regions representing the uncertainty in the FRB position relative to its host galaxy nucleus.

Following the scheme in Mannings et al. (2021), we apply a weighted 2D Gaussian probability distribution centered on the FRB localization ellipse to determine the mean and variance in $\theta$. Because the host of FRB 20210117A is not detected, it is not included in this analysis. FRB 20181112A is also not included due to (i) the large uncertainty relative to the size of the host and (ii) the host not being identified by GALIGHT at sufficient significance (see Section 3.1).

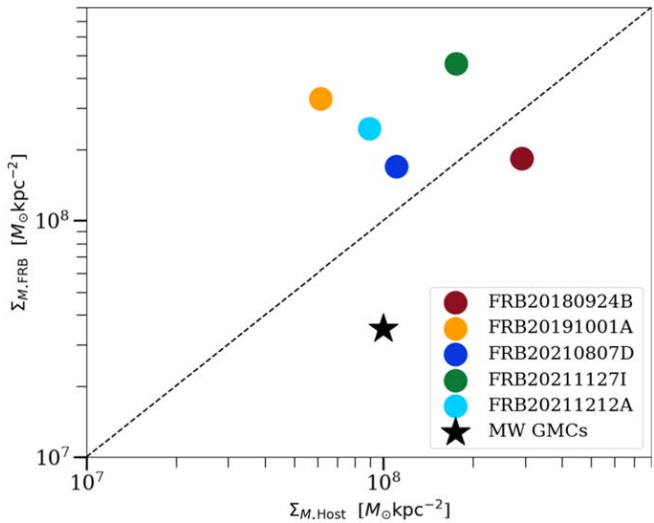

**Figure 3.** The average stellar mass surface density at the burst site vs. average stellar mass surface density of the host. The dashed line shows the 1:1 relation. The error bars for each point are significantly smaller than the marker size. For context, we have also plotted giant molecular clouds (GMCs; black star) within the Milky Way. We note that the FRBs preferentially occur in regions of higher local surface density than the Milky Way GMCs.

The projected angular offsets ($\theta$) range between 0″7 and 5″0, with a median value of 2″0. These values are included in Table 4. We then converted these angular offsets to physical offsets using the host redshifts and our assumed cosmological model (Section 1). The physical offsets range from 2.0 to 11.9 kpc, with a median value of 3.4 kpc. For comparison, we determine the median of the distribution of Galactic globular clusters presented in Baumgardt & Vasiliev (2021), which is 7.7 kpc. Finally, we used the half-light radii $r_e$ (see Section 3.1) of the hosts to determine the host-normalized offsets $\delta R/r_e$. The median host-normalized offset for this sample is 2.1 $r_e$. This is larger than the median expected offset if FRBs traced the locations of stars in their disks and fell within the half-light radius of the host galaxy (i.e., 1 $r_e$).

To determine the uncertainty on the cumulative distributions, we again utilize the method of Heintz et al. (2020) as described in Section 3.4. In Figure 5, the left panel shows the physical offset CDF for the combined sample, and the right panel shows the host-normalized CDF for the combined sample of FRBs. The gray shaded region depicts the uncertainty for each distribution.

Again, we utilize the KS test, with a $P_{KS} < 0.05$ requirement in order to reject the null hypothesis. We test against CCSNe (Schulze et al. 2021), SGRBs and LGRBs (Fong et al. 2010; Fong & Berger 2013; Blanchard et al. 2016), Ca-rich transients (Lunnan et al. 2017; De et al. 2020), superluminous supernovae (SLSNe; Lunnan et al. 2015; Schulze et al. 2021), and Type Ia SNe (Uddin et al. 2020). We also test against a distribution of magnetars as modeled in Safarzadeh et al. (2020). The resulting $P_{KS}$ values are shown in the legend of Figure 5.

We find that, with the resulting $P_{KS}$ values, the KS test rejects all transient samples except CCSNe for projected offsets. In the case of host-normalized offsets, only Ca-rich transients are rejected.

### 4.2. Offset from Spiral Arms

To investigate the location of FRBs relative to the spiral arms of their host galaxies, we determined the minimum distance from the FRB to the closest projected spiral arm in the host galaxy. This spiral arm offset is referred to as SO_min.

To estimate SO_min, we first convolve the GALIGHT residual image (the original image minus the GALIGHT model; see Section 3.1) using a smoothing kernel of 5 pixels. We then mask all values below a threshold of 0.6—chosen after experimentation—to isolate pixels not associated with spiral arms. Custom masks were then created for each host to mask sources clearly not associated with the spiral arms of the host (e.g., nearby galaxies, or field stars in projection). All unmasked pixels were taken as part of the host galaxy spiral arm structure. We then calculated the minimum 2D Gaussian-weighted offset between the FRB and the nearest unmasked pixel. This methodology mirrors that of galactocentric offset calculation. These SO_min values are reported in Table 4. We show an illustrative image of this analysis in Figure 6, where we define $0.25\sigma$ contours as the edges of the spirals.

The SO_min values range from 0″07 to 0″55 with a median value of 0″42 or physical minimum separations of 0.11–2.36 kpc with a median value of 0.53 kpc. Two of the five FRBs (20210807D and 20211212A) have an SO_min value consistent with zero at 95% confidence.

## 5. Discussion

### 5.1. Using AO to Image FRB Host Galaxies

This data set presents the first ground-based AO sample of FRB host galaxies. This demonstrates our ability to match space-based depth and resolution with ground-based resources in the near-IR. Given the increasing sample of precisely localized FRBs, it will be critical to include ground-based observations to complement more expensive, space-based surveys.

While ground-based observations are currently limited to IR wavelengths, these do provide a detailed look at the distributions of stellar mass in these hosts, morphologies, and the connections between bursts and older stellar populations. The $5\sigma$ magnitude limits, indicating the deepest objects detectable in these images, are listed in Table 3. In comparison with the limiting magnitude calculations obtained with HST in Mannings et al. (2021), we have achieved a similar depth using GeMS/GSAOI with a fraction of the integration time (between 500 and 2000 s) in comparison to almost 3000 s with HST, thanks to the light-gathering power of an 8 m telescope. With HST in Mannings et al. (2021), the spatial resolution is ∼0″2, while our GeMS/GSAOI images had a delivered image resolution of ∼0″06.

### 5.2. Spiral Structure in FRB Hosts

Four out of five hosts in the GeMS/GSAOI sample that were successfully imaged show clear spiral arm structure even though these galaxies were not preselected based on known spiral arm morphology. These FRBs are all nonrepeaters. The host galaxies are at a range of inclination angles from 33° (FRB 20211127I) to 59° (FRB 20210807D).

Using GALIGHT, we determined the half-light radii of these five host galaxies and produced residual images which further highlight the spiral and low surface brightness structures in

Table 4
Offsets and Light Locations of FRBs

| FRB | $\theta$ (arcsec) | $\delta R$ (kpc) | $\delta R/r_e$ | $SO_{min}$ (arcsec) | $SO_{min}$ (kpc) | $F_F$ |
|---|---|---|---|---|---|---|
| 20180924B | 0.72 ± 0.17 | 3.47 ± 0.83 | 8.09 ± 1.93 | 0.49 ± 0.07 | 2.36 ± 0.34 | 0.55 ± 0.26 |
| 20191001A | 0.94 ± 0.18 | 3.63 ± 0.69 | 2.09 ± 0.40 | 0.32 ± 0.12 | 1.21 ± 0.46 | 0.64 ± 0.11 |
| 20210807D | 5.00 ± 0.40 | 11.91 ± 0.95 | 4.32 ± 0.34 | 0.07 ± 0.07 | 0.16 ± 0.16 | 0.47 ± 0.24 |
| 20211127I | 2.18 ± 0.40 | 2.07 ± 0.38 | 0.74 ± 0.13 | 0.55 ± 0.28 | 0.53 ± 0.27 | 0.24 ± 0.10 |
| 20211212A | 2.04 ± 0.72 | 2.84 ± 1.00 | 1.20 ± 0.42 | 0.08 ± 0.09 | 0.11 ± 0.12 | 0.67 ± 0.18 |
| Median | 2.04 ± 0.72 | 3.47 ± 0.83 | 2.09 ± 0.40 | 0.32 ± 0.12 | 0.53 ± 0.27 | 0.55 ± 0.26 |

Note. $\theta$ refers to the angular offset of the FRB from the center of the host galaxy. $\delta R$ is this offset in physical units. $\delta R/r_e$ is the host-normalized offset. $SO_{min}$ is the offset of the FRB from the closest spiral arm of the host galaxy. $F_F$ is fractional flux.

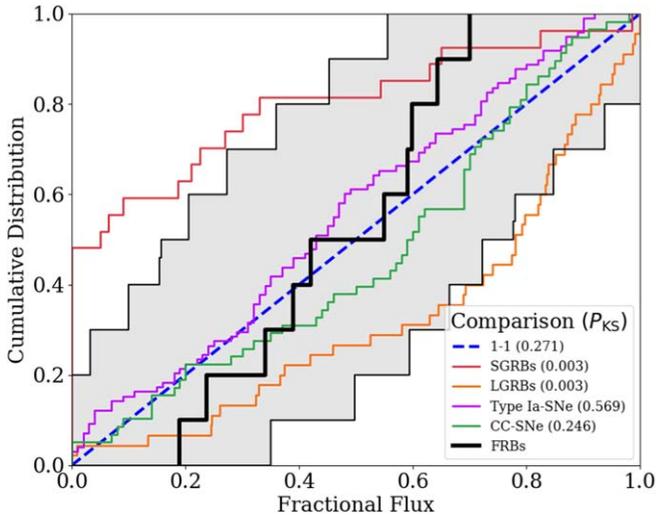

Figure 4. Cumulative distribution of the fractional flux of the combined FRB sample (ten FRBs): the five FRBs in the sample that had hosts detected in our $K_s$ AO imaging, in addition to the FRBs from Mannings et al. (2021) that were imaged with F160W [1385.77–1700.31 nm]. The gray shaded region is a bootstrap estimate of the rms of the distribution, which accounts for both uncertainties in individual measurements and statistical uncertainties due to the sample size. We compare this distribution to those of SGRBs (red; Fong et al. 2010; Fong & Berger 2013), LGRBs (orange; Blanchard et al. 2016), Type Ia SNe (purple; Wang et al. 2013), CCSNe (green; Lunnan et al. 2015), and a 1-1 line shown in blue. The SGRB and LGRB samples are rejected by the KS test, implying they are not from the same underlying population as FRBs.

these hosts. Echoing the results of Mannings et al. (2021), we show that most of the bursts that occur in spiral hosts also occur on or near the spiral structures. This proximity to regions of relatively concentrated stellar mass is also illustrated by the $\Sigma_{M_*\mathrm{FRB}}$ (see Figure 3), which shows more than half of the FRBs in the GeMS/GSAOI sample occurring in a region of elevated stellar mass density. This is in stark contrast to SGRBs, for example, as they occur at high offsets and are therefore separated from the stellar mass in their hosts (e.g., Fong et al. 2022).

Using the residual images, we determine the minimum distance of each FRB to the closest spiral arm in the host galaxy (designated $SO_{min}$). The $SO_{min}$ values range from 0.11 to 2.36 kpc with a median value of 0.53 kpc. The FRBs are significantly closer to spiral arms of their host compared to the center of their host galaxy, with a median galactocentric offset being 3.47 kpc. The FRBs tend to be on or near a spiral arm but not coincident with the brightest portions of the spiral arm. We discuss the light at the location of the FRB in more detail in Section 5.3.

Abdeen et al. (2022) find evidence supporting the existence of age gradients in spiral arms, as predicted by density wave theory (Lin et al. 1969). Stars between 0 and 10 Myr tend to closely map the spiral arms of the host galaxy, trailing the arm slightly (see Figures 3 and 4 of Abdeen et al. 2022). Magnetars are a compelling theoretical source of FRBs due to their compact size and extreme magnetizations. Notably, they are included in this younger population of stars. Therefore, if a majority of FRBs are associated with young magnetars, we would expect these to occur on or slightly trailing a spiral arm. FRBs 20210807D and 20211212A, within their respective uncertainties, are potentially coincident with a spiral arm of their host. FRBs 20191001A and 20211127I could potentially be leading one spiral arm or trailing another.

Aramyan et al. (2016) and Karapetyan (2022) map the locations of extremely well-localized Type II and Type Ia SNe, respectively, relative to the spiral arms of their host galaxies. The authors created radial light profiles from the center of the galaxy extending through the transient's location. They were able to precisely determine the location of each SN within the spiral arm of the respective host galaxy. Aramyan et al. (2016) find that each SN in their sample is within the inner and outer bounds of a spiral arm, which are defined by a zero threshold in the residual galaxy image. This residual image is found by subtracting a bulge+disk model from the original image, similar to what we have done in this study. However, there are variations according to each type of SN. For example, Type Ibc SNe occur closer to the leading edges of the arms than do Type II SNe. Karapetyan (2022) finds that the "normal" Type Ia SNe (defined by the light-curve decline rate) span a range of distances from the spiral arm peaks, occurring within the edges of the spiral arms as well as in interarm regions. In the future, with a move toward (i) consistent milliarcsecond scale localizations and (ii) larger sample sizes due to CHIME Outriggers, CRACO on ASKAP, and DSA-110, we hope to expand the study of spiral arm offsets in FRBs to obtain more precise measurements of the location of FRBs relative to the substructure of the spiral arms of their hosts.

### 5.3. FRB Galactocentric Offsets and Light Locations

We calculated the projected offset of the FRBs from their host centers. We combined this GeMS/GSAOI sample of FRBs with the FRBs from Mannings et al. (2021) to create Figure 5 with a larger sample of FRBs (ten FRBs). At high statistical significance, five out of six transient classes included (SGRBs, LGRBs, Ca-rich transients, and SLSNe) are rejected as coming from the same underlying population as FRBs. We are unable to reject this null hypothesis with respect to CCSNe and Type Ia SNe. In Heintz et al. (2020), CCSNe are confidently not rejected.

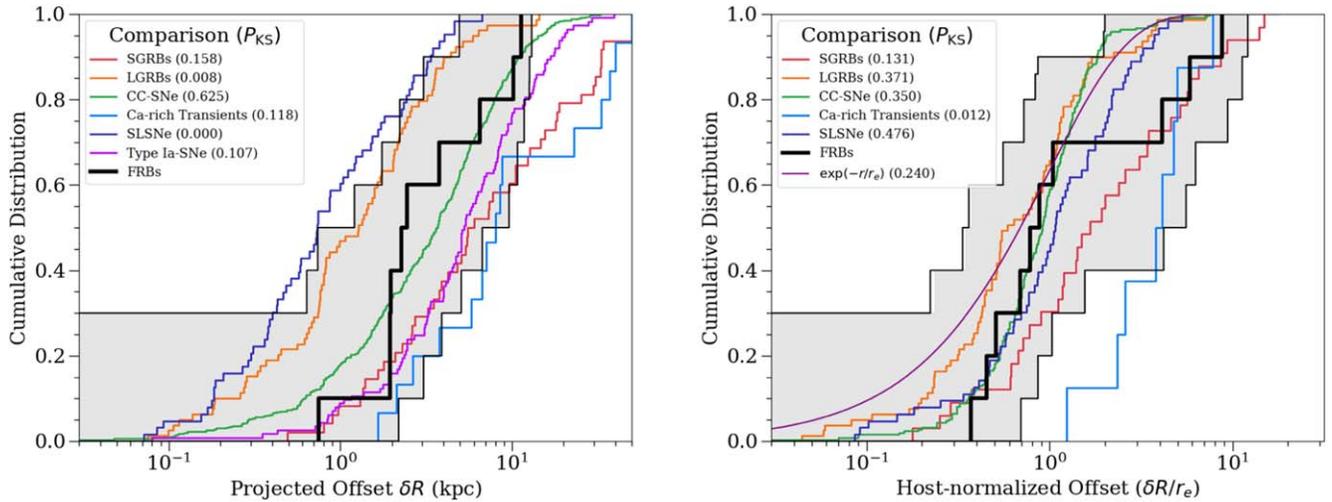

**Figure 5.** Left: cumulative distribution of the offsets of the combined FRB sample (ten FRBs). We plot other transient classes for comparison including SGRBs (red; Fong et al. 2010; Fong & Berger 2013), LGRBs (orange; Blanchard et al. 2016), Ca-rich transients (light blue; Lunnan et al. 2017; De et al. 2020), CCSNe (green; Schulze et al. 2021), SLSNe (dark blue; Lunnan 2015; Schulze et al. 2021; Hsu et al. 2023), and Type Ia SNe (purple; Uddin et al. 2020). Right: same as left but using host-normalized offset.

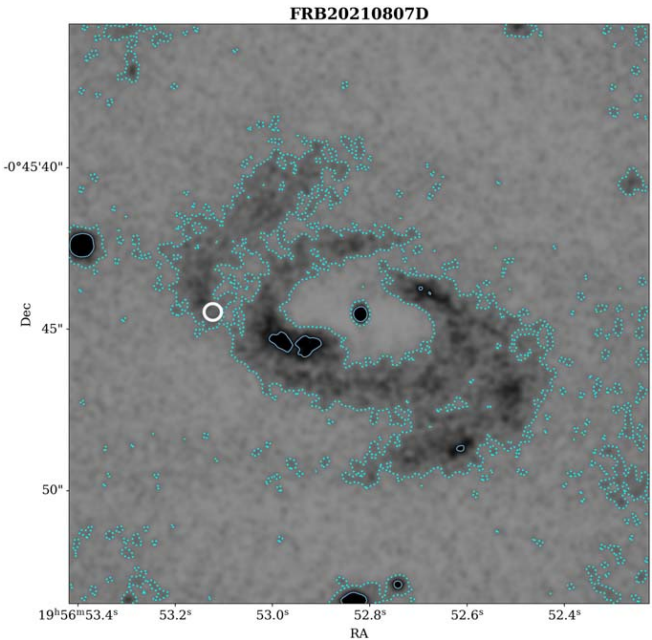

**Figure 6.** Example contour plot showing the extent of the spiral structure in the host of FRB 20210807D. The $SO_{min}$ computes the distance between the FRB localization (shown in white) and the nearest unmasked pixel that is determined to be part of the spiral arm. The contours highlight the $0.25\sigma$ and $1\sigma$ levels. The $SO_{min}$ measured for this burst is $0\rlap{.}''07$.

We use the effective radius from GALIGHT to calculate the host-normalized projected offset of each FRB. Through a KS test shown in the right panel of Figure 5, Ca-rich transients are rejected as coming from the same underlying population as FRBs, while SGRBs, LGRBs, and CCSNe are not rejected. Similarly, Heintz et al. (2020) do not reject CCSNe.

In terms of fractional flux, FRBs are distinctly different from LGRBs, which tend to occur in brighter regions of their host galaxies (e.g., Fruchter et al. 2006). The locations of FRBs with respect to their host's brightness are also distinct from SGRBs (e.g., Fong et al. 2010). FRBs, however, are not distinguished from CCSNe and Type Ia SNe in terms of fractional flux.

As shown in Figures 4 and 5, the hypothesis that FRBs and CCSNe are from the same underlying population is not rejected based on projected offset, host-normalized offset, or fractional flux. Our findings on the local environment scale are consistent with the findings of Bochenek et al. (2021), which investigated global host properties. In Bochenek et al. (2021), they compared a sample of FRB host galaxies with the galaxies of CCSNe, LGRBs, and SLSNe. LGRBs and SLSNe were rejected as coming from the same underlying population as FRBs, while CCSNe were not rejected. The Milky Way magnetar population is dominated by magnetars born in CCSNe (Bochenek et al. 2021). Thus, if magnetars are the dominant FRB progenitor, and the Milky Way magnetar population is representative of the magnetar population overall, then FRB host galaxies should be consistent with the host galaxies of CCSNe (Bochenek et al. 2021). However, while this connection appears plausible for the sample of FRBs presented in this paper, we cannot assume the Milky Way magnetar population is representative of all FRB progenitors. There are several exceptions to the typical spiral host galaxy, such as the FRB from a globular cluster (e.g., Bhardwaj et al. 2021), FRBs in dwarf galaxies (e.g., Bhandari et al. 2023), and various FRBs that are not coincident with star-forming regions in their host galaxies (e.g., Dong et al. 2023), which is where we might expect more massive progenitors to be located. Additionally, in these tests, we assume that the FRBs in this sample all have the same physical origin, which is not necessarily the case. The possible magnetar connection—or more clearly, CCSN connection—seems relevant for this sample of FRBs specifically, but broad characteristics of FRB host galaxies overall might point toward a multitude of pathways. The possibility for multiple pathways is supported by the ability to explain some of the exceptions listed above (FRB in a globular cluster, elliptical hosts, etc.) with a Type Ia SNe related source or progenitor. In order to differentiate between possible connections to these different transient classes, we must at least double the sample size.

The confidence in these *individual* measurements is largely limited by the FRB localization precision. The pixel scale of GSAOI in particular is approximately $0\rlap{.}''02$ per pixel. Not until we reach a precision of 10 s of milliarcsecond localizations will the measurements be limited by image resolution. Even still, future studies that leverage the higher-precision localizations will be able to glean a great deal of information about FRB local environments with existing high-resolution imaging resources and the next generation of large telescopes.

## 6. Conclusions

In this paper, we use high-resolution images from GeMS/GSAOI to perform a detailed study of the environments of five FRBs. Using these data, we make measurements of galactocentric offset (Section 4), fractional flux (Section 3.4), stellar mass surface density (Section 3.2), and offset from spiral arms (Section 4.2). We also demonstrate the feasibility of creating samples of high-resolution imaging of FRB hosts using AO-supported ground-based resources.

From each of these measures, we found that FRBs are located at moderate offsets (median host-normalized offset of 2.09 $r_e$), compared to other transient classes such as LGRBs and Ca-rich transients. These bursts also appear on or close to the spiral structure in their hosts, with a median distance of 0.53 kpc.

As for the light at the FRB location, the IR fractional flux demonstrated that the combined samples of FRBs trace the distribution of stellar mass, similar to CCSNe. We also examined the FRB location in terms of stellar mass surface density, finding four out of five FRB sites to be elevated in this quantity compared to the global value. In combination, these findings point toward a possible similarity between FRB local environments and those of CCSNe. This is a finding supported by the conclusions of Bochenek et al. (2021), which finds similarities between the global properties of FRBs hosts and CCSNe hosts. However, we must take into account the inability to reject an underlying connection between FRBs and Type Ia SNe. The local environment and host characteristics should be explored with respect to Type Ia SNe, in general, and different Type Ia progenitors, specifically (as explored in Karapetyan 2022).

With many well-localized FRB detections looming on the horizon due to upgraded experiments and new instruments coming online, we will be able to better characterize the properties of FRB environments relative to their host galaxies and other transient classes. Doubling the number of FRB detections with localizations comparable to those presented here will already provide greater opportunity for discernment. With a 5× or 10× increase in sample size, along with milliarcsecond precision localizations, it seems we will be able to make strong conclusions about FRB local environments and origins. Additionally, the increasing sensitivities of instruments will facilitate studies at higher redshifts. It is necessary that we utilize all resources available to prepare for and attend to the large number of incoming FRB detections and precise localizations. In combination with possible James Webb Space Telescope, HST, future large telescopes, and archival data, AO will prove to be an effective tool.

## Acknowledgments

Based on observations obtained at the international Gemini Observatory, a program of NSF's NOIRLab, which is managed by the Association of Universities for Research in Astronomy (AURA) under a cooperative agreement with the National Science Foundation on behalf of the Gemini Observatory partnership: the National Science Foundation (United States), National Research Council (Canada), Agencia Nacional de Investigación y Desarrollo (Chile), Ministerio de Ciencia, Tecnología e Innovación (Argentina), Ministério da Ciência, Tecnologia, Inovações e Comunicações (Brazil), and Korea Astronomy and Space Science Institute (Republic of Korea).

M.N.W. and R.A.J. gratefully acknowledge support from NSF grant AST-2206492 and from the Nantucket Maria Mitchell Association. Authors M.N.W., A.G.M., J.X.P., S.S, R.A.J., and N.T., as members of the Fast and Fortunate for FRB Follow-up team, acknowledge support from NSF grants AST-1911140, AST-1910471, and AST-2206490. A.T.D. acknowledges support through Australian Research Council Discovery Project DP220102305. C.D.K. acknowledges partial support from a CIERA postdoctoral fellowship. M.G. is supported by the Australian Government through the Australian Research Council's Discovery Projects funding scheme (DP210102103).

## ORCID iDs

Michele N. Woodland https://orcid.org/0009-0001-3334-9482
J. Xavier Prochaska https://orcid.org/0000-0002-7738-6875
Stuart D. Ryder https://orcid.org/0000-0003-4501-8100
Lachlan Marnoch https://orcid.org/0000-0003-1483-0147
Regina A. Jorgenson https://orcid.org/0000-0003-2973-0472
Sunil Simha https://orcid.org/0000-0003-3801-1496
Nicolas Tejos https://orcid.org/0000-0002-1883-4252
Alexa Gordon https://orcid.org/0000-0002-5025-4645
Wen-fai Fong https://orcid.org/0000-0002-7374-935X
Charles D. Kilpatrick https://orcid.org/0000-0002-5740-7747
Adam T. Deller https://orcid.org/0000-0001-9434-3837
Marcin Glowacki https://orcid.org/0000-0002-5067-8894

## References

Abbott, T. M. C., Adamów, M., Aguena, M., et al. 2021, ApJS, 255, 20
Abdeen, S., Davis, B. L., Eufrasio, R., et al. 2022, MNRAS, 512, 366
Aggarwal, K., Budavári, T., Deller, A. T., et al. 2021, ApJ, 911, 95
Aramyan, L. S., Hakobyan, A. A., Petrosian, A. R., et al. 2016, MNRAS, 459, 3130
Bannister, K. W., Deller, A. T., Phillips, C., et al. 2019, Sci, 365, 565
Bannister, K. W., Shannon, R. M., Macquart, J. P., et al. 2017, ApJL, 841, L12
Bassa, C. G., Tendulkar, S. P., Adams, E. A. K., et al. 2017, ApJL, 843, L8
Baumgardt, H., & Vasiliev, E. 2021, MNRAS, 505, 5957
Bhandari, S., Bannister, K. W., Lenc, E., et al. 2020b, ApJL, 901, L20
Bhandari, S., Gordon, A. C., Scott, D. R., et al. 2023, ApJ, 948, 67
Bhandari, S., Sadler, E. M., Prochaska, J. X., et al. 2020a, ApJL, 895, L37
Bhardwaj, M., Gaensler, B. M., Kaspi, V. M., et al. 2021, ApJL, 910, L18
Blanchard, P. K., Berger, E., & Fong, W.-f. 2016, ApJ, 817, 144
Bochenek, C. D., Ravi, V., Belov, K. V., et al. 2020, Natur, 587, 59
Bochenek, C. D., Ravi, V., & Dong, D. 2021, ApJL, 907, L31
Chambers, K. C., Magnier, E. A., Metcalfe, N., et al. 2016, arXiv:1612.05560
Chatterjee, S., Law, C. J., Wharton, R. S., et al. 2017, Natur, 541, 58
CHIME/FRB Collaboration, Amiri, M., Andersen, B. C., et al. 2021, ApJS, 257, 59
CHIME/FRB Collaboration, Andersen, B. C., Bandura, K. M., et al. 2020, Natur, 587, 54
Chittidi, J. S., Simha, S., Mannings, A., et al. 2021, ApJ, 922, 173

De, K., Kasliwal, M. M., Tzanidakis, A., et al. 2020, ApJ, 905, 58
Ding, X., Birrer, S., Treu, T., & Silverman, J. D. 2021, arXiv:2111.08721
Dong, Y., Eftekhari, T., Fong, W.-f., et al. 2024, ApJ, 961, 15
Driessen, L. N., Barr, E., Buckley, D., et al. 2024, MNRAS, 527, 3659
Flewelling, H. A., Magnier, E. A., Chambers, K. C., et al. 2020, ApJS, 251, 7
Fong, W., & Berger, E. 2013, ApJ, 776, 18
Fong, W., Berger, E., Blanchard, P. K., et al. 2017, ApJL, 848, L23
Fong, W., Berger, E., & Fox, D. B. 2010, ApJ, 708, 9
Fong, W.-f., Nugent, A. E., Dong, Y., et al. 2022, ApJ, 940, 56
Fruchter, A. S., Levan, A. J., Strolger, L., et al. 2006, Natur, 441, 463
Glowacki, M., Lee-Waddell, K., Deller, A. T., et al. 2023, ApJ, 949, 25
Gordon, A. C., Fong, W.-f., Kilpatrick, C. D., et al. 2023, ApJ, 954, 40
Heintz, K. E., Prochaska, J. X., Simha, S., et al. 2020, ApJ, 903, 152
Hsu, B., Blanchard, P. K., Berger, E., & Gomez, S. 2024, ApJ, 961, 24
Karapetyan, A. G. 2022, MNRAS, 517, L132
Kirsten, F., Marcote, B., Nimmo, K., et al. 2022, Natur, 602, 585
Kocz, J., Ravi, V., Catha, M., et al. 2019, MNRAS, 489, 919
Kuncarayakti, H., Anderson, J. P., Galbany, L., et al. 2018, A&A, 613, A35
Labrie, K., Anderson, K., Cárdenes, R., Simpson, C., & Turner, J. E. H. 2019,
    in ASP Conf. Ser. 523, Astronomical Data Analysis Software and Systems
    XXVII, ed. P. J. Teuben et al. (San Francisco, CA: ASP), 321
Lada, C. J., & Dame, T. M. 2020, ApJ, 898, 3
Law, C. J., Sharma, K., Ravi, V., et al. 2024, ApJ, 967, 18
Leung, C., Mena-Parra, J., Masui, K., et al. 2021, AJ, 161, 81
Lin, C. C., Yuan, C., & Shu, F. H. 1969, ApJ, 155, 721
Lorimer, D. R., Bailes, M., McLaughlin, M. A., Narkevic, D. J., &
    Crawford, F. 2007, Sci, 318, 777
Lunnan, R., Chornock, R., Berger, E., et al. 2015, ApJ, 804, 90
Lunnan, R., Kasliwal, M. M., Cao, Y., et al. 2017, ApJ, 836, 60
Mannings, A. G., Fong, W.-f., Simha, S., et al. 2021, ApJ, 917, 75
Marcote, B., Nimmo, K., Hessels, J. W. T., et al. 2020, Natur, 577, 190
Marcote, B., Paragi, Z., Hessels, J. W. T., et al. 2017, ApJL, 834, L8
Niu, C. H., Aggarwal, K., Li, D., et al. 2022, Natur, 606, 873
Planck Collaboration, Ade, P. A. R., Aghanim, N., et al. 2016, A&A, 594, A13
Prochaska, J. X., Neeleman, M., Kanekar, N., & Rafelski, M. 2019, ApJL,
    886, L35
Ravi, V., Catha, M., Chen, G., et al. 2023, ApJL, 949, L3
Ryder, S. D., Bannister, K. W., Bhandari, S., et al. 2023, Sci, 382, 294
Safarzadeh, M., Prochaska, J. X., Heintz, K. E., & Fong, W.-f. 2020, ApJL,
    905, L30
Schulze, S., Yaron, O., Sollerman, J., et al. 2021, ApJS, 255, 29
Tendulkar, S. P., Bassa, C. G., Cordes, J. M., et al. 2017, ApJL, 834, L7
Tendulkar, S. P., Gil de Paz, A., Kirichenko, A. Y., et al. 2021, ApJL, 908, L12
Uddin, S. A., Burns, C. R., Phillips, M. M., et al. 2020, ApJ, 901, 143
Wang, X., Wang, L., Filippenko, A. V., Zhang, T., & Zhao, X. 2013, Sci,
    340, 170
Xu, H., Niu, J. R., Chen, P., et al. 2022, Natur, 609, 685
Zhang, B. B., Zhang, B., Sun, H., et al. 2018, NatCo, 9, 447